\documentclass[runningheads]{llncs}

\usepackage{amssymb}
\usepackage[english]{babel}
\usepackage[cp1250]{inputenc}
\usepackage{graphicx}
\usepackage[T1]{fontenc}
\usepackage{url}

\newcommand{\keywords}[1]{\par\addvspace\baselineskip
\noindent\keywordname\enspace\ignorespaces#1}

\begin{document}

\mainmatter  % start of an individual contribution

% first the title is needed
\title{The Inhabitation Problem for Rank Two Intersection Types \footnote{Partly suported by KBN grant 3~T11C~002~27}}
\titlerunning{The Inhabitation Problem for Rank Two Intersection Types}
\author{Dariusz Kuśmierek}

\institute{Warsaw University, Institute of Informatics\\
Banacha 2, 02-097 Warsaw, Poland\\
\email{\tt{daku@mimuw.edu.pl}}
}

\toctitle{The Inhabitation Problem for Rank Two Intersection Types}
\maketitle

\begin{abstract}
We prove that the inhabitation problem for rank two intersection types is decidable, but (contrary to a common belief) \linebreak \rm{EXPTIME}-hard. The exponential time hardness is~shown by~reduction from the~in-place acceptance problem for~alternating Turing machines.
\end{abstract}
\keywords{lambda calculus, intersection types, type inhabitation problem, alternating Turing machine}

\newtheorem{tw}{Theorem}%[section]
\newtheorem{lem}[tw]{Lemma}
\newtheorem{df}[tw]{Definition}
\newtheorem{fact}[tw]{Fact}
\newtheorem{wniosek}[tw]{Corollary}

%\tableofcontents
%\listoffigures
%\listoftables

\subsection*{Introduction}
\addcontentsline{toc}{subsection}{Introduction}
Type inhabitation problem is~usually defined as~follows: ``does there exist a closed term $T$ of a given type $\tau$ (in~an~empty environment)''. The~answer to~this question depends on~the~system and~its~type inference rules.
By~the Curry-Howard correspondence, the inhabitation problem for a given type system is equivalent to the 
decidability of the corresponding logic.

In the simply typed system the inhabitation problem is \rm{PSPACE}-complete (see [7]).

The intersection types system studied in the current paper allows types of the form $\alpha \cap \beta$. 
Intuitively, a term can be assigned the type $\alpha \cap \beta$ if and only if it can be assigned the type $\alpha$ and also the type $\beta$. 
This system corresponds to the logic of ``strong conjunction'' (see [1, 4, 5]). 

Undecidability of~the general inhabitation problem for intersection types was shown by P. Urzyczyn in~[8].

Several weakened systems were studied, and proved to be decidable.
T. Kurata and M. Takahashi in~[2] proved the~decidability of~the~problem in the system $\lambda(E \cap, \leq)$ which does not use the~rule~$(I \cap)$. 

Leivant in [3] defines the rank of an intersection type.
The notion of rank turns out to be very useful, since it provides means for classification and a~measure of complexity of the intersection types. 

One can notice, that construction in [8] uses only types of rank four. 
The inhabitation for rank three is still a well-known open problem. The problem for rank two was so far believed to be decidable in polynomial space (see [8]).

Our result contradicts this belief. We prove the inhabitation problem for rank two to be \rm{EXPTIME}-hard by a reduction from the halting problem for Alternating Linear Bounded Automata (ALBA in short).
The idea of the reduction is as follows. For a given ALBA and a given word of length $n$ we construct a type of the form $\alpha_1 \cap \ldots \cap \alpha_n \cap \alpha_{n+1} \cap \alpha_{n+2}$. The intended meaning of the components $\alpha_1$, $\ldots$, $\alpha_n$ is that $\alpha_i$ represents the behaviour of the $i$-th cell of the tape, the $\alpha_{n+1}$ represents changes in the position of the head of machine, and the last part $\alpha_{n+2}$ simulates changes of the machine state. The $\cap$ operator is used here to hold and process information about the whole configuration of the automata.

The fact that the problem for rank two is \rm{EXPTIME}-hard only highlights how difficult the still open problem for rank three can be.

%\begin{center}
%\begin{tabular}{cc}
%$(\leq)$ & $\frac{\displaystyle{\Gamma \vdash M \colon \alpha \quad \alpha \leq \beta}} {\displaystyle{ \Gamma %\vdash M\colon \beta}}$ \\
 
%\end{tabular}
%\end{center}

\section{Basics}\label{r:pojecia}
\subsection{Intersection Types}
We consider a~lambda calculus with types defined by the following induction:
\begin{itemize}
\item Type variables are types
\item If $\alpha$ and $\beta$ are types, then $\alpha \rightarrow \beta$ and $\alpha \cap \beta$ are also types
\end{itemize}
We assume that the operator $\cap$ is associative, commutative and idempotent. That is, the meaning of the intersection type of the form $\alpha_1 \cap \ldots \cap \alpha_m$ does not depend on the order or number of occurences of each of the types $\alpha_i$.

The type inference rules for our system are as follows:\\

\begin{tabular}{lllr}

(VAR) & $\Gamma \vdash x \colon \sigma$ & & if $(x \colon \sigma) \in \Gamma$ \\
 & & \\
 
$(E \rightarrow)$ & $\frac{\displaystyle{\vphantom{\frac{a}{a}}\Gamma \vdash M \colon \alpha \rightarrow \beta \quad \Gamma \vdash N \colon \alpha}} {\displaystyle{\vphantom{\frac{a}{a}}\Gamma \vdash (MN) \colon \beta}}$ & \\
 & & &\\

$(I \rightarrow)$ & $\frac{\displaystyle{\vphantom{\frac{a}{a}}\Gamma, (x \colon \alpha) \vdash M \colon \beta}}{\displaystyle{\vphantom{\frac{a}{a}}\Gamma \vdash \lambda x.M \colon \alpha \rightarrow \beta}}$ & \\
 & & & \\

$(E \cap)$ & $\frac{\displaystyle{\vphantom{\frac{a}{a}}\Gamma \vdash M \colon \alpha \cap \beta}}{\displaystyle{\vphantom{\frac{a}{a}}\Gamma \vdash M \colon \alpha}}$ & & $\frac{\displaystyle{\vphantom{\frac{a}{a}}\Gamma \vdash M \colon \alpha \cap \beta}}{\displaystyle{\vphantom{\frac{a}{a}}\Gamma \vdash M \colon \beta}}$\\
 & & & \\

$(I \cap)$ &  $\frac{\displaystyle{\vphantom{\frac{a}{a}}\Gamma \vdash M \colon \alpha \quad \Gamma \vdash M \colon \beta}}{\displaystyle{\vphantom{\frac{a}{a}}\Gamma \vdash M \colon \alpha \cap \beta}}$ & & \\

\end{tabular}
\\ \\
By~the Curry-Howard correspondence the calculus defined above corresponds to~the~logic of ``strong conjunction'' of~Mints and~Lopez-Escobar (see [1, 4, 5]). While considering the rule~$(I \cap)$ one~can~easily notice the~specific binding of the types $\alpha$~and~$\beta$ by~the~term $M$. A term~has a~type $\alpha \cap \beta$ if~it~has a type $\alpha$ and at~the~same time it~has a type $\beta$. Accordingly in the corresponding logic, one~can prove a~formula $\alpha \cap \beta$ if one can prove $\alpha$~and~$\beta$ and both~formulae have at~least one~common proof (strong conjunction). It follows that~checking the correctness of~formulae in~the~logic of ``strong conjunction'' may prove to~be conceptually difficult and~usually it is more convenient to~regard the~equivalent problem of~inhabitation for~the intersection type systems.

\subsection{Intersection Types Classification}
\begin{df}\rm 
Following Leivant ([3]) we define the \emph{rank} of a type $\tau$, denoted by $rank(\tau)$:
\begin{center}
\begin{tabular}{l}
$rank(\tau) = 0$, if $\tau$ is a simple type (without ``$\cap$'');\\
$rank(\tau \cap \sigma) = max(1, rank(\tau), rank(\sigma))$;   \\
$rank(\tau \rightarrow \sigma) = max(1+rank(\tau), rank(\sigma))$, when $rank(\tau)>0$ or $rank(\sigma)>0$.\\
\end{tabular}
\end{center}
\end{df}

\section{Decidability of the Inhabitation Problem}\label{r:rozstrzygalnosc}
\subsection{The Algorithm}

\begin{df}\rm
A variable $x$ is \textit{$k$-ary} in an environment $\Gamma$, if $\Gamma$ includes a declaration $(x \colon \alpha)$, such that
\begin{center}
$\alpha = \beta_1 \to \cdots \to \beta_k \to \tau$\\ 
or\\ 
$\alpha = \gamma \cap (\beta_1 \to \cdots \to \beta_k \to \tau)$
\end{center}
\end{df}
The variable is~$k$-ary, if~one can~apply it~to~some $k$~arguments of~certain types.

\paragraph{The Algorithm.}\label{sec:Dowod1}
The task considered by the algorithm in each step is a set of problems:

\begin{center}
$[\Gamma_1 \vdash  M\colon\tau_1 $, $\ldots$, $ \Gamma_n \vdash  M\colon\tau_n]$\\
($n$ environments, $n$ types, but only one term to be found)
\end{center}
Our algorithm uses the ``intersection removal'' operation $Rem$ defined as follows:
\begin{center}
\begin{tabular}{l}
$Rem(\Gamma \vdash  M\colon\tau) = \{\Gamma \vdash  M\colon\tau\}  $ if $\tau$ is not an intersection\\ 
$Rem(\Gamma \vdash  M\colon\tau_1 \cap \tau_2) =  Rem(\Gamma \vdash  M\colon\tau_1) \cup Rem(\Gamma \vdash  M\colon\tau_2)$\\
\end{tabular} 
\end{center}
The purpose for the $Rem$ operation is to eliminate ``$\cap$'' and to convert a judgement $\Gamma \vdash  M\colon\tau$ with $\tau$ being possibly an intersection type into a set of judgements where the types on the right side are not intersections.

For a given type $\tau$ the first task is:

\begin{center}
$Z_0 = Rem(\emptyset \vdash  M\colon\tau)$\\
\end{center}
It is worth noting that in the tasks processed by our procedure types on the right side will never be intersections.

Let the current task be: 		
\begin{center}
$Z = [\Gamma_1 \vdash  M\colon\tau_1 $, $\ldots$, $ \Gamma_n \vdash  M\colon\tau_n]$\\ 
\end{center}

\begin{enumerate}
\item If each type $\tau_i$ is of the form $\alpha_i \to \beta_i$, 
then the next task processed recursively by the algorithm will be:
\begin{center}
$Z' = Rem(\Gamma_1 \cup \{ x \colon \alpha_1 \} \vdash  M'\colon\beta_1) \cup \ldots \cup Rem(\Gamma_n \cup \{ x \colon \alpha_n \} \vdash  M'\colon\beta_n)$,\\ 
\end{center}
where $x$ is a fresh variable not used in any of the $\Gamma_i$.
			
If the recursive call for $Z'$ returns $M'$, then $M=\lambda x.M'$, if on the other hand the recursive call gives an answer ``empty type'', we shall give the same answer. 
	\item If at least one of the $\tau_i$ is a type variable, then the term $M$ cannot be a an abstraction, but must be an application or a variable. Suppose that there exists a variable $x$ which is $k$-ary in each of the environments, and for each~$i$ it holds that:
\begin{center}
	$\Gamma_i \vdash x \colon \beta_{i1} \to \cdots \to \beta_{ik} \to \tau_i$\\ 
\end{center}
(if there is more than one such variable, we pick nondeterministically one of them). Then:
	
			\begin{itemize}
				\item If $k=0$, then $M=x$,
				\item If $k>0$, then $M=xM_1 \ldots M_k$,
				where $M_i$ are solutions for the $k$ independent problems:
					\begin{center}
						$Z_1 = [\Gamma_1 \vdash  M_1\colon\beta_{11} $, $\ldots$, $ \Gamma_n \vdash  M_1\colon\beta_{n1}]$,\\
						\ldots\\
						$Z_k = [\Gamma_1 \vdash  M_k\colon\beta_{1k} $, $\ldots$, $ \Gamma_n \vdash  M_k\colon\beta_{nk}]$\\
					\end{center}
				If any of the $k$ recursive cals gives the answer ``empty type'', we shall give the same answer. 
			\end{itemize}
	\item Otherwise we give the answer ``empty type''. \\

	\end{enumerate}

\begin{fact}
If the above algorithm finds a term $M$ for an input type $\tau$, then $\vdash M \colon \tau$.
\end{fact}
\paragraph{Proof.}
Straightforward.

\subsection{Soundness}
The algorithm described above is not capable of finding all the terms of a given type. Hence, to prove the corectness of the proposed procedure, we first define the notion of a \textit {long} solution, then we show that every task, which has a solution, has also a long solution. Finally we complete the proof of the soundness of the algorithm by proving that every long solution can be found by the given procedure.

\begin{df}\rm
$M$ is a \textit{long} solution of the task $Z=[\Gamma_1 \vdash  M\colon \tau_1 , \ldots,\Gamma_n \vdash  M\colon \tau_n]$, when one of the following holds:
\begin{itemize}
\item All types $\tau_i$ are of the form $\alpha_i \rightarrow \beta_i$ and $M=\lambda x.M'$, where $M'$ is a~\rm{long} solution of the task $Z' = Rem([\Gamma_1, (x \colon \alpha_1) \vdash  M\colon \beta_1 , \ldots,\Gamma_n, (x \colon \alpha_n) \vdash  M\colon \beta_n])$, or
\item Some $\tau_i$ is a type variable and $M=xM_1 \ldots M_k$, where for $i = 1 \ldots n$ \linebreak $\Gamma_i \vdash x \colon \alpha_{i1} \rightarrow \cdots \rightarrow \alpha_{ik} \rightarrow \tau_i$ and $M_1, \ldots, M_k$ are \rm{long} solutions of tasks $Z_1, \ldots, Z_k$, where $Z_j = [\Gamma_1 \vdash  M_j\colon \alpha_{1j} , \ldots,\Gamma_n \vdash  M_j\colon \alpha_{nj}]$ for $j= 1 \ldots k$.
\end{itemize} 
\end{df}

\begin{lem}\label{lem:posiada}
If there exists a solution of a task $Z$, then there exists a long one.
\end{lem}
\paragraph{Proof.}
Assume that $M$ is a solution of $Z=[\Gamma_1 \vdash  M\colon \tau_1 , \ldots,\Gamma_n \vdash  M\colon \tau_n]$. We~construct a long solution $A(M,Z)$ in the following way:
\begin{itemize}
\item If there is a $\tau_i$ which is a type variable, then $M$ is not an abstraction and:\\
	\begin{itemize}
		\item If $M=x$, then $A(M,Z) = M$, because in this case $M$ is a long solution,
		\item If $M=xM_1 \ldots M_k$, then it must hold that $\Gamma_i \vdash x \colon \alpha_{i1} \rightarrow \cdots \rightarrow \alpha_{ik}~\rightarrow~\tau_i$ for $i~=~1~\ldots~n$, so $A(M,Z) = xA(M_1,Z_1) \ldots A(M_k,Z_k)$, where $Z_j = [\Gamma_1 \vdash  M_j\colon \alpha_{1j} , \ldots,\Gamma_n \vdash  M_j\colon \alpha_{nj}]$ for $j= 1 \ldots k$.\\		
	\end{itemize}
\item Otherwise (if all $\tau_i$ have the form of $\alpha_i \rightarrow \beta_i$):\\
	\begin{itemize}
		\item If $M=xM_1 \ldots M_k$ (possibly for $k=0$), then $A(M,Z) = \lambda z.A(Mz,Z')$, where $Z'=[\Gamma_1, (z \colon \alpha_1) \vdash  Mz\colon \beta_1 , \ldots,\Gamma_n, (z \colon \alpha_n) \vdash  Mz\colon \beta_n]$ (since $Mz$ is a solution of the task $Z'$, and $\lambda z.Mz$ -- of the task $Z$).
		\item If $M=\lambda x.M'$, then $A(M,Z) = \lambda x.A(M',Z')$, where \linebreak $Z'=Rem([\Gamma_1, (x \colon \alpha_1) \vdash  M'\colon \beta_1 , \ldots, \Gamma_n, (x \colon \alpha_n) \vdash  M'\colon \beta_n])$.\\
	\end{itemize}

\end{itemize}
\begin{lem}\label{lem:kazde_dlugie}
Every long solution $M$ of the task $Z=[\Gamma_1 \vdash  M\colon \tau_1 , \ldots,\Gamma_n \vdash  M\colon \tau_n]$ can be found by the above alternating procedure.
\end{lem}
\paragraph{Proof.}
By induction on the structure of $M$.
\begin{itemize}
	\item $M=x$. Since $M$ is long, at least one of the $\tau_i$ must be a type variable. Hence the algorithm working on the task $Z$ will search in the environments $\Gamma_i$ for a variable of the right type (case 2 of the algorithm). One of the variables that the algorithm chooses from is of course $x$.
	\item $M=xM_1 \ldots M_k$. Like before we can reason that the algorithm shall choose the case 2, and in one of its possible runs the algorithm will choose the variable $x$. After $x$ is chosen, the procedure shall search for solutions of the tasks $Z_1, \ldots, Z_k$. By the definition of a long solution, we have that $M_1, \ldots, M_k$ are long solutions of the tasks $Z_1, \ldots, Z_k$, and by the induction hypothesis, these solutions can be found by the recursive runs of our procedure. It follows that also $M$ can be found.
	\item $M=\lambda x.M'$. Then of course all the types $\tau_i$ have to be of the form $\alpha_i \rightarrow \beta_i$. Hence for the task $Z$ the procedure shall choose case 1, and search for solution of the task $Z'=Rem([\Gamma_1, (x \colon \alpha_1) \vdash  M'\colon \beta_1 , \ldots, \Gamma_n, (x \colon \alpha_n) \vdash  M'\colon \beta_n])$. By the induction hypothesis, a long solution $M'$ for $Z'$ can be found by the algorithm.
\end{itemize}

\begin{wniosek}
Our algorithm finds an inhabitant for every non-empty type, for which it terminates. 
\end{wniosek}
\paragraph{Proof.}
A direct conclusion of Lemmas \ref{lem:posiada} and \ref{lem:kazde_dlugie}.
\section{The Termination of the Algorithm}\label{r:algorytm}

Let us consider the work of the algorithm for a type $\tau$ of rank two.

\begin{fact}
Types of variables put in the environments during the work of algorithm are of the rank at most one. 
\end{fact}
\paragraph{Proof.}
The environments are modified only in case 1. If any of the variables put in the enviroments was of rank two, than the type $\tau$ must have been of the rank three. 

\begin{fact}
In every recursive run there is no task with more than $|\tau|$ simultaneous problems to solve. 
\end{fact}
\paragraph{Proof.}
When a new problem is generated by the $Rem$ operation, one ``$\cap$'' is removed from the type $\tau$ and the type is split between the problems. There can never be more than $|\tau|$ problems. The recursive calls in case 2 do not create any new parallel problems, because of the way they are created. Namely, in these problems the procedure searches for terms which can serve as arguments for a~variable taken from the environment. As we noticed before, in environments there are only variables with types of rank zero and one, and such variables can only be given arguments with types of rank zero. And these types are simple (without intersections), so they do not generate new problems by the $Rem$ operation.

\subsection{The Decidability}

\begin{tw}\label{tw:latwosc}
The inhabitation problem for rank two intersection types is decidable.
\end{tw}
\paragraph{Proof.}
First notice that the environments cannot grow bigger infinitely during the work of the algorithm. 
Variables are added to the environments only when all currently examined types $\tau_i$ are of the form $\alpha_i \to \beta_i$. Then every environment $\Gamma_i$ is expanded by a new variable of the type $\alpha_i$. Note that there are only $O(|\tau|)$ types that can be assigned to a variable in one environment. Since we do not need to keep several variables of the same type (meaning of the same type in each of the environments) it follows that there is finite number of possible distinct environments that may occur during the work of the algorithm. Also the number of the types that may occur on the right hand side of each $\vdash$ is $O(|\tau|)$. Hence each branch of the alternating procedure must finish or repeat a~configuration in a~finite (although possibly exponential) number of steps.

\section{The Lower Bound}\label{r:problem}

\subsection{Terms of Exponential Size}\label{sec:przyklad}
First we shall consider an instructive example. We propose a schema for creating instances of the inhabiation problem for which the above algorithm has to perform an exponential number of steps before finding the only inhabitant. The size of the inhabitant will also be exponential in the size of the type. Our example demonstrates a technique used in the construction to follow.\linebreak Let $T(n) = \tau_{1} \cap \ldots \cap \tau_{n}$, where

\begin{center}
$
\tau_{i} = 
\alpha 
\rightarrow 
	\underbrace{		
	 \Psi
		\rightarrow
		\cdots
			\rightarrow \Psi
	}_{i-1}
\rightarrow
	(\alpha \rightarrow \beta)
\rightarrow 
	\underbrace{
	 	(\beta \rightarrow \alpha)
	 	\cdots
	 	\rightarrow 
	 	(\beta \rightarrow \alpha)
	}_{n-i}	
\rightarrow 
	\beta
$, and\\
$\Psi=(\alpha \rightarrow \alpha) \cap (\beta \rightarrow \beta)$.
\end{center}
For instance $T(3) = $
\begin{center}

\begin{tabular} {cccccccccc}
	$(\alpha $ & $	\rightarrow$ & 	$(\alpha \rightarrow \beta) $ & $\rightarrow$ &  $(\beta \rightarrow \alpha)	$ & $\rightarrow $ & $	(\beta \rightarrow \alpha)	$ & $\rightarrow $ & $	\beta)$ & $\cap$\\
	$(\alpha 	$ & $\rightarrow$ & 	$\Psi $ & $\rightarrow$ &  $(\alpha \rightarrow \beta)	$ & $\rightarrow $ & $	(\beta \rightarrow \alpha)	$ & $\rightarrow $ & $	\beta)$ & $\cap$\\
	$(\alpha 	$ & $\rightarrow$ & 	$\Psi $ & $\rightarrow$ &  $\Psi $ & $	\rightarrow $ & $	(\alpha \rightarrow \beta)	$ & $ \rightarrow $ & $	\beta) $ &\\

\end{tabular}
\end{center}
One can notice that a construction of an inhabitant for this type is similar to the rewriting process from the word $\beta\beta\beta$ to the word $\alpha\alpha\alpha$, and it is a letter by letter rewritng. For $|T(n)| = O(n^2)$, there is only one term $t$ of type $T(n)$, and $|t| = O(2^n)$. For instance, the only (modulo $\alpha$--equivalence) term of type $T(3)$ is:
\begin{center}
	$\lambda x_1x_2x_3x_4.x_2(x_3(x_2(x_4(x_2(x_3(x_2x_1))))))$.
\end{center}
While for $T(4)$ it is:
\begin{center}
	$\lambda x_1x_2x_3x_4x_5.x_2(x_3(x_2(x_4(x_2(x_3(x_2(x_5(x_2(x_3(x_2(x_4(x_2(x_3(x_2x_1))))))))))))))$.
\end{center}
In what follows, while proving \rm{EXPTIME}-hardness of the inhabitation problem, we shall generate types of a similar form to $T(n)$. For this reason it is worth to use $T(n)$ for introducing notions and notations, which we shall use later on. \\
Because of the different role played by the ``$\cap$'' and ``$\rightarrow$'' it is convenient to consider the structure of the type in terms of columns and rows. The rows are connected with ``$\cap$'', and columns with ``$\rightarrow$'' (in the case of $T(n)$ there are $n$ rows: $\tau_1$, $\ldots$, $\tau_n$). According to this terminology $T(3)$ has three rows and five columns. One row represents operations available for a given object and the initial and final state of the object (here states are variables $\alpha$~and~$\beta$). One column represents a certain operation (that is a step of a certain automaton). In type $T(3)$ there are three available operations. The $i$-th operation changes the $i$-th sign from $\beta$~to~$\alpha$, and all earlier signs from $\alpha$ to $\beta$. More precisely each $(\alpha \rightarrow \beta)$ in the type $T(n)$ represents the change from $\beta$ to $\alpha$, and an occurence of $\Psi$ represents no change of sign (changes $\beta$ and $\alpha$ to themselves).

\subsection{EXPTIME-hardness}
We shall show the lower bound for the complexity of the inhabitation problem by a reduction from the \rm{EXPTIME}-complete problem of the~in-place acceptance for~alternating Turing machines.\\
\begin{df}\rm 
\emph{An alternating Turing machine} is a quintuple:
\begin{center}
$M = (Q, \Gamma, \delta, q_0, g)$, where
\end{center}
\begin{itemize}
	\item $Q$ is a non-empty, finite set of states.
	\item $\Gamma$ is a non-empty, finite set of symbols. We shall assume that $\Gamma = \left\{0,1\right\}$.
	\item $\delta \subseteq (Q\times\Gamma) \times (Q\times\Gamma\times\left\{L,R\right\})$ - is a non-empty, finite transition relation.
	\item $q_0 \in Q$ is the initial state.
	\item $g: Q \rightarrow \left\{\wedge, \vee, accept\right\}$ is a function which assigns a kind of every state.
\end{itemize}
\end{df}

\begin{df}\rm 
A \emph{configuration} of an alternating Turing machine machine is a~triple:
\begin{center}
$C = (q, t, n)$, where 
\end{center}
\begin{itemize}
	\item $q \in Q$ is a state
	\item $t \in \Gamma^*$ is a tape content
	\item $n \in N$ is a position of the head
\end{itemize}
\end{df}

\begin{df}\rm 
An \emph{Alternating Linear Bounded Automaton} (ALBA) is an alternating Turing machine whose tape head never leaves the input word.
\end{df}

\begin{df}\rm 
We shall say that a \emph{transition} $p=((q_1,s_1),(q_2,s_2,k))$ is \emph{consistent} with the relation $\delta$ in a configuration $C_1 = (q_1,t,n)$, when the following conditions hold:
\begin{itemize}
	\item $p \in \delta$; 
	\item $t(n)=s_1$;
	\item ($k=L$ and $n>1$) or ($k=R$ and $n<|t|$).
\end{itemize}
We shall say then that $p$ \emph{transforms} a configuration $C_1$ to a configuration $C_2~=~(q_2, t_2, n_2)$, where
\begin{itemize}
	\item $
t_2(m)=\left\{
	\begin{array}{ll}
 s_2 & \textrm{if $m=n$}\\
 t(m) & \textrm{otherwise}
 \end{array}\right. 
$
	\item $
n_2=\left\{
	\begin{array}{ll}
 n+1 & \textrm{if $k=R$}\\
 n-1 & \textrm{otherwise}
 \end{array}\right. 
$
\end{itemize}
\end{df}

\begin{df}\rm 
An alternating Turing machine \emph{accepts} in configuration \linebreak $C~=~(q,t,n)$, if
\begin{itemize}
	\item $g(q) = accept$ and $|t| = n$, or
	\item $g(q) = \vee$ and there exists a transition consistent with $\delta$, which transforms the configuration $C$ to a configuration in which the automaton accepts, or
	\item $g(q) = \wedge$ and every transition consistent with $\delta$, transforms $C$ to a configuration in which the automaton accepts.
\end{itemize}
\end{df}

It is worth noting, that in configurations in which the head scans the first symbol of the tape, the only available transitions are these, which move the head to the right, and when the head reaches the end of the word, the only active transitions will move it to the left.

\begin{df}\rm 
\emph{The problem of in-place acceptance} is defined as follows: does a~given ALBA accept a given word $t$ (meaning it accepts the configuration $C_0~=~(q_0,t,1)$). 
\end{df}

It is known that \rm{APSPACE} = \rm{EXPTIME} (see Corollary 2 to Theorem 16.5 and Corollary 3 to Theorem 20.2 in [6]).

\begin{lem}
The problem of in-place acceptance for ALBA is \rm{EXPTIME}-com\-plete (\rm{APSPACE}-complete).
\end{lem}
\paragraph{Proof.}
A simple modification of the proof of Theorem 19.9 in [6]. 
First we note that the in-place acceptance is in \rm{APSPACE}. Consider a machine \linebreak $M~=~(Q, \Gamma, \delta, q_0, g)$.
Keeping the counter of steps, we simulate the run of~$M$ on the input word~$t$. We reject if $M$ rejects, or if machine makes more than $|t||Q||\Gamma|^{|t|}$ steps, because after so many steps machine has to repeat a configuration.\\
Let $L$ be a language in \rm{APSPACE} accepted in space $n^k$ by a machine~$M$. It means that $M$ does not use in any of its parallel computations more than $n^k$ cells of the tape (where $n$ is the length of the input word). Let us denote the blank symbol by $\bot$. Let us consider a modified machine $M'$, which during its work performs the same moves as $M$, but when $M$ reaches an accepting state, the head of $M'$ makes $n^k-n$ steps to the right and also enters an accepting state. It is clear that $M$ accepts $t$ if and only if the machine $M'$ accepts $t \bot^{n^k-n}$ without ever leaving this word (note that according to the definition, machine $M$ accepts with the head at rightmost symbol of the input word) --- blank symbols $\bot$ at the very end of the word do not change the behaviour of the machine, and $M$ does not use more than $n^k$ cells of the tape. So $t$ belongs to $L$ if and only if $M'$ accepts $t \bot^{n^k-n}$ in-place.

\begin{tw}\label{tw:trudnosc}
The inhabitation problem for rank two intersection types is \linebreak \rm{EXPTIME}-hard.
\end{tw}
\paragraph{Proof.}
Let us consider the input word $t = t_1t_2 \ldots t_{n-1}t_n$. We construct a type with $n+2$ rows and some number of columns (according to the terminology introduced in \ref{sec:przyklad}). The first $n$ rows shall represent the state of $n$ tape cells. The next row shall represent the position of the head (values $1 \ldots n$). The last row shall stand for the state of the machine. Let $q_{acc}$ be a new type variable. We shall begin our construction with these two columns:

\begin{center}
\begin{tabular}{rcccccl}

$($ & $\ldots$ & $\rightarrow$ & $2$  & $\rightarrow$ & $t_1$ & $)\cap$ \\

 & & & $\ldots$ & & & \\
$($ & $\ldots$ & $\rightarrow$ & $2$  & $\rightarrow$ & $t_n$ & $)\cap$ \\
$($ & $\ldots$ & $\rightarrow$ & $0$  & $\rightarrow$ & $1$ & $)\cap$ \\
$($ & $\ldots$ & $\rightarrow$ & $q_{acc}$  & $\rightarrow$ & $q_0$ & $)$ \\

\end{tabular}
\end{center}
where $q_{acc} \notin Q$. The last column in the type represents the initial configuration: in the cells there are symbols from the input word $t$, the variables $t_1 \ldots t_n$ represent the input word, the head is at first position, and the machine is in state~$q_0$. The second last column represents the final state. \\
\indent In the further construction we shall add new columns on the left side.

\subsubsection{Accepting States:}\label{stany_acc}
If the machine has any accepting states, we add a column responsible for a transition from the accepting states to our additional state $q_{acc}$. Let the $q_1 \ldots q_r$ be all the accepting states of the machine. The additional column will be:

\begin{center}
\begin{tabular}{l}
$
\left. 
	\begin{array}{l}
 S \rightarrow \\

 \ldots \\
 S \rightarrow \\ 
 \end{array} \right\}n
$ \\
$	\begin{array}{l}
 K \rightarrow \\

 Q \rightarrow \\ 
 \end{array}
$
%$ K \rightarrow $\\
%$ Q \rightarrow $ \\

\end{tabular}
\end{center} where\\
\begin{center}
$S = ( 2 \rightarrow 0 ) \cap (2 \rightarrow 1)$, \\
$K = (0 \rightarrow n)$, \\
$Q = ( q_{acc} \rightarrow q_1 ) \cap \ldots \cap (q_{acc} \rightarrow q_r)$. \\

\end{center}
Each column of the type (except the last one) will be assigned to one variable in a term. The components of a column are just different types that are assigned to the same variable in $n+2$ different environments.
The variable, which will have assigned types being parts of this column, is responsible for transition from each accepting state of the machine (for each tape content and for head of machine being at last sign of the word) to the state $q_{acc}$.

\subsubsection{States of Kind $\vee$:}\label{stany_lub}

Let $Id(p)=(\overbrace{0 \rightarrow \cdots \rightarrow 0}^{p+1})\cap 
(\overbrace{1 \rightarrow \cdots \rightarrow 1}^{p+1})$. For each element $((q_1,s_1),(q_2,s_2,k))$ of~$\delta$, such that $g(q_1) = \vee$, we add $n-1$ columns --- one column for each position of the head.\\ 
If $k=L$, then the $i$-th added column is of the form:

\begin{center}
\begin{tabular}{l}
$
\left.
	\begin{array}{l}
 Id(1) \rightarrow \\
 \ldots \\
 Id(1) \rightarrow\\ 
 \end{array} \right\}i
$ \\
$ (s_2 \rightarrow s_1) \rightarrow $ \\
$
\left. 
	\begin{array}{l}
 Id(1) \rightarrow \\
 \ldots \\
 Id(1) \rightarrow\\ 
 \end{array} \right\}n-i-1
$ \\
$ (i-1 \rightarrow i) \rightarrow $ \\
$ (q_2 \rightarrow q_1) \rightarrow $ \\

\end{tabular}
\end{center}
\pagebreak[3]
And if $k=R$, then the $i$-th added column is:

\begin{center}
\begin{tabular}{l}

$
\left.
	\begin{array}{l}
 Id(1) \rightarrow \\
 \ldots \\
 Id(1) \rightarrow\\ 
 \end{array} \right\}i-1
$ \\
$ (s_2 \rightarrow s_1) \rightarrow $ \\
$
\left. 
	\begin{array}{l}
 Id(1) \rightarrow \\
 \ldots \\
 Id(1) \rightarrow\\ 
 \end{array} \right\}n-i
$ \\
$ (i+1 \rightarrow i) \rightarrow $ \\
$ (q_2 \rightarrow q_1) \rightarrow $ \\

\end{tabular}
\end{center}

\subsubsection{States of Kind $\wedge$:}\label{stany_i}
For each state $q$, such that $g(q) = \wedge$, and for each sign $s \in \Gamma$ we add $n$ columns (one for each position of the head). The $i$-th column is generated this way:
let $((q,s),(q_1,s_1,k_1)), \ldots, ((q,s),(q_p,s_p,k_p))$ be all transitions available in $q$, when head is at $i$-th position, which holds sign $s$. In this case, the $i$-th column has the form of:

\begin{center}
\begin{tabular}{l}
$
\left.
	\begin{array}{l}
 Id(p) \rightarrow \\
 \ldots \\
 Id(p) \rightarrow\\ 
 \end{array} \right\}i-1
$ \\
$ (s_1 \rightarrow \cdots \rightarrow s_p \rightarrow s) \rightarrow $ \\
$
\left. 
	\begin{array}{l}
 Id(p) \rightarrow \\
 \ldots \\
 Id(p) \rightarrow\\ 
 \end{array} \right\}n-i
$ \\
$ ((i+r(k_1)) \rightarrow \cdots \rightarrow (i+r(k_p)) \rightarrow i) \rightarrow $ \\
$ (q_1 \rightarrow \cdots \rightarrow q_p \rightarrow q) \rightarrow $ \\

\end{tabular}
\end{center}
where
\begin{center}
$
r(k)=\left\{
	\begin{array}{ll}
 1 & \textrm{if $k=R$}\\
 -1 & \textrm{otherwise}
 \end{array}\right. 
$
\end{center}
The above construction corresponds to the definition of an acceptance in a state of the kind $\wedge$, when the automaton needs to accept in all the reachable configurations. The variable corresponding to the added column can be used in term (inhabitant) only when it is possible to find inhabitants for each of the arguments. Each such inhabitant represents a computation in one of the possible configurations (after executing the apropriate step). \\
Note that, if there is no reachable configuration from a state of the kind $\wedge$, then the added column will not be of a functional type (it will not have any arrows except for the one on the right), and so it will not require any further searching for inhabitants. The computation will terminate successfully, which corresponds to the acceptation of a word in states of kind $\wedge$, from which the machine has nowhere to go.

\subsection{Correctness of the Reduction}
We shall consider the instances of the type inhabitation problem generated by the above construction. Notice that, for such types, the construction of the inhabitant according to the algorithm proposed in section \ref{r:rozstrzygalnosc} will go as follows: first the problem shall be split into $n+2$ subproblems by use of the $Rem$ operator, then the algorithm will use the case 1 serveral times, after which the current task will be 
\begin{center}
$Z=[\Gamma_1 \vdash  T\colon s_1 , \ldots, \Gamma_n \vdash  T\colon s_n, \Gamma_{n+1}~\vdash T\colon k, \Gamma_{n+2} \vdash T\colon q]$. 
\end{center}
From this moment algorithm shall use only the application case (case 2), since the types under consideration shall always be type variables. In the following steps the only thing that shall change will be $s_1, \ldots, s_n, k, q$, but the environments $\Gamma_1, \ldots, \Gamma_n$ shall stay the same.   
 
\begin{lem}
Let $Z=[\Gamma_1 \vdash  T\colon s_1 , \ldots, \Gamma_n \vdash  T\colon s_n, \Gamma_{n+1} \vdash T\colon k, \Gamma_{n+2} \vdash T\colon q]$ be a~task for a type generated by the above construction for an alternating machine~$M$.
For $q \in Q$ the task $Z$ has a solution if and only if the machine $M$ accepts in place the configuration $C=(q,s_1\ldots s_n,k)$.
\end{lem}
\paragraph{Proof.} 
$ \\(\Rightarrow)$ Induction with respect to the structure of the solution $T$ of the task $Z$. \\

	\begin{itemize}
		\item $T$ is a variable $x$. Then $\Gamma_{n+2} \vdash x \colon q$. Then $q$ is of the kind ``$\wedge$'', and from the configuration $C$ there are no transitions consistent with $\delta$ (because only in this case there was a variable of a type being a type variable added to the environments (see \ref{stany_i})). Hence $M$~accepts the configuration $C$.
		\item $T$ is an abstraction. Impossible, because the types, for which we seek an inhabitant in $Z$ are type variables.
		\item $T$ is an application. There are three possibilities.
		\begin{itemize}
				\item $T = x_1x_{acc}$, where $\Gamma_{n+2} \vdash x_1 \colon q_{acc} \rightarrow q$. According to the type construction (see \ref{stany_acc}), the only variables, which in the environment $\Gamma_{n+2}$ can be supplied with the argument of the type $q_{acc}$ are variables representing accepting states. Hence $q$ is an accepting state of $M$, so $M$ accepts in $C$.
		\item $T = xT_1$ and $g(q) = \vee$. According to the type construction (see \ref{stany_lub}) it holds that:
\begin{center} 
$\Gamma_1 \vdash  x\colon s_1 \rightarrow s_1,$\\
$\ldots$\\
$\Gamma_k \vdash x \colon s_k' \rightarrow s_k,$\\
$\ldots$\\
$\Gamma_n \vdash  x\colon s_n \rightarrow s_n,$\\
$\Gamma_{n+1} \vdash x\colon k+r(c) \rightarrow k,$\\
$\Gamma_{n+2} \vdash x\colon q' \rightarrow q$, \\
\end{center}

and $((q,s_k),(q',s_k', c)) \in \delta$. Hence $T_1$ is a solution of the task \\

$[\Gamma_1 \vdash  T_1\colon s_1,$ $\ldots,$ $\Gamma_k \vdash T_1 \colon s_k',$ $\ldots,$ $\Gamma_n \vdash  T_1\colon
 s_n,$
\begin{flushright}
$\Gamma_{n+1} \vdash T_1\colon k+r(c),$ $\Gamma_{n+2} \vdash T_1\colon q']$. 
\end{flushright} $ $

By the induction hypothesis (for $T_1$) the machine $M$ accepts in \linebreak $C_1~=~(q', s_1 \ldots s_k' \ldots s_n, k~+~r(c))$. However, since $q$ is of the kind~$\vee$ and there exists a transition from $C$ do $C_1$, it follows that $M$ accepts also $C$.
		
		\item $T = xT_1 \ldots T_m$, for some $m$ and $g(q) = \wedge$. According to the type construction (see \ref{stany_i}) it must hold that: 
\begin{center}
$\Gamma_1 \vdash  x\colon s_1 \rightarrow \cdots \rightarrow s_1 \rightarrow s_1,$\\
$\ldots$\\
$\Gamma_k \vdash x \colon s_{k1} \rightarrow \cdots \rightarrow s_{km} \rightarrow s_k,$\\
$\ldots$\\
$\Gamma_n \vdash  x\colon s_n \rightarrow \cdots \rightarrow s_n \rightarrow s_n,$ \\
$\Gamma_{n+1} \vdash x\colon k+r(c_1) \rightarrow \cdots \rightarrow k+r(c_m) \rightarrow k,$ \\
$\Gamma_{n+2} \vdash x\colon q_1 \rightarrow \cdots \rightarrow q_m \rightarrow q$
\end{center}
 and the following transitions are all consistent with $\delta$ transitions from $C$: $((q,s_k),(q_1, s_{k1}, c_1)), \ldots, ((q,s_k),(q_m, s_{km}, c_m))$. Then of course each $T_i$ is a solution of the task \\

 $[\Gamma_1 \vdash  T_i\colon s_1,$ $\ldots,$ $ \Gamma_k \vdash T_i \colon s_{ki},$ $\ldots,$ $\Gamma_n \vdash  T_i\colon s_n,$
\begin{flushright}
 $\Gamma_{n+1} \vdash T_i\colon k+r(c_i),$ $\Gamma_{n+2} \vdash T_i\colon q_i]$.
\end{flushright} $ $

By the induction hypothesis for $T_1, \ldots, T_m$, the machine $M$ accepts in all of the $C_1, \ldots, C_m$, where $C_i = (q_i, s_1 \ldots s_{ki} \ldots s_n, k+r(c_i))$. It means that $M$ accepts in all configurations reachable from $C$, so it accepts in~$C$.		
		\end{itemize}
	\end{itemize}
		
\pagebreak[5]
$\\(\Leftarrow)$  Induction with respect to the definition of acceptance. \\ \\
	(\textit{Base}) Let $g(q) = accept$. Then $k=n$, because the machine accepts only with the head in the rightmost position. Then according to the construction for accepting types (see \ref{stany_acc}), there exists a variable $x$, such that: $\Gamma_1 \vdash  x\colon 2 \rightarrow~s_1,$ $\ldots$, $\Gamma_n \vdash  x\colon 2 \rightarrow s_n, \Gamma_{n+1} \vdash x\colon 0 \rightarrow n, \Gamma_{n+2} \vdash x\colon q_{acc} \rightarrow q$. So $T=xx_{acc}$ is a solution of $Z$.\\ \\
	(\textit{Step}) Assume that $M$ accepts in $C= (q, s_1 \ldots s_n, k)$, where $q$ is not an accepting state. There are two possibilities:
	\begin{itemize}
		\item Let $g(q) = \vee$. Since $M$ accepts in the configuration $C$ it means that there exists a transition $((q,s_k), (q',s_k',c))$, such that $M$ accepts in configuration $C_1 = (q', s_1 \ldots s_k' \ldots s_n, k+r(c))$. According to the induction hypothesis there exists a solution $T_1$ of the task \\
		
$[\Gamma_1 \vdash  T_1\colon s_1,$ $\ldots,$ $\Gamma_k \vdash T_1 \colon s_k',$ $\ldots,$ $\Gamma_n \vdash  T_1\colon s_n,$
\begin{flushright}
$\Gamma_{n+1} \vdash T_1\colon k+r(c),$ $\Gamma_{n+2} \vdash T_1\colon q']$. 
\end{flushright} $ $

Since $q$ is of the kind $\vee$, then according to the construction (see \ref{stany_lub}) there exists a variable $x$, such that 
\begin{center}
$\Gamma_1 \vdash  x\colon s_1 \rightarrow s_1,$\\
$\ldots$\\
$\Gamma_k \vdash x \colon s_k' \rightarrow s_k,$\\
$\ldots$\\
$\Gamma_n \vdash  x\colon s_n \rightarrow s_n,$\\
$\Gamma_{n+1} \vdash x\colon k+r(c) \rightarrow k,$\\
$\Gamma_{n+2} \vdash x\colon q' \rightarrow q$. 
\end{center}
So $T = xT_1$ is a solution of the task $Z$.
		\item $g(q) = \wedge$. Then for each transition $((q,s_k), (q_i, s_{ki}, c_i))$ available from $C$, machine $M$ accepts in configuration $C_i = (q_i, s_1 \ldots s_{ki} \ldots s_n, k+r(c_i))$. By the induction hypothesis $T_1, \ldots, T_m$ are solutions of the tasks $Z_1, \ldots, Z_m$,~where \\
		
$Z_i = [\Gamma_1 \vdash  T_i\colon s_1,$ $\ldots,$ $\Gamma_k \vdash T_i \colon s_{ki},$ $\ldots,$ $\Gamma_n \vdash  T_i\colon s_n,$
\begin{flushright}
$\Gamma_{n+1} \vdash T_i\colon k+r(c_i),$ $\Gamma_{n+2}~\vdash~T_i\colon~q_i]$.
\end{flushright} $ $

Since $q$ is of the kind $\wedge$, there must (see \ref{stany_i}) exist a variable $x$, such that 
\begin{center}
$\Gamma_1 \vdash  x\colon s_1 \rightarrow \cdots \rightarrow s_1 \rightarrow s_1,$\\
$\ldots$\\
$\Gamma_k \vdash x \colon s_{k1} \rightarrow \cdots \rightarrow s_{km} \rightarrow s_k,$\\
$\ldots$\\
$\Gamma_n \vdash  x\colon s_n \rightarrow \cdots \rightarrow s_n \rightarrow s_n,$\\
$\Gamma_{n+1} \vdash x\colon k+r(c_1) \rightarrow \cdots \rightarrow k+r(c_m) \rightarrow k,$\\
$\Gamma_{n+2} \vdash x\colon q_1 \rightarrow \cdots \rightarrow q_m \rightarrow q$.
\end{center}
 Then $T = xT_1 \ldots T_m$ is a solution of $Z$.
	\end{itemize}

\end{document}